\documentstyle[natbib2,natbibmnfix,epsfig]{mn}
\bibpunct[,]{(}{)}{;}{a}{}{}

\newif\ifAMStwofonts

\ifoldfss
  \ifCUPmtlplainloaded \else
    \NewTextAlphabet{textbfit} {cmbxti10} {}
    \NewTextAlphabet{textbfss} {cmssbx10} {}
    \NewMathAlphabet{mathbfit} {cmbxti10} {} 
    \NewMathAlphabet{mathbfss} {cmssbx10} {} 
  \fi
  \ifAMStwofonts
    \ifCUPmtlplainloaded \else
      \NewSymbolFont{upmath} {eurm10}
      \NewSymbolFont{AMSa} {msam10}
      \NewMathSymbol{\upi}     {0}{upmath}{19}
      \NewMathSymbol{\umu}     {0}{upmath}{16}
      \NewMathSymbol{\upartial}{0}{upmath}{40}
      \NewMathSymbol{\leqslant}{3}{AMSa}{36}
      \NewMathSymbol{\geqslant}{3}{AMSa}{3E}

       \let\le=\leqslant
       
    \fi
  \fi
\fi 

\ifnfssone
  \newmathalphabet{\mathit}
  \addtoversion{normal}{\mathit}{cmr}{m}{it}
  \addtoversion{bold}{\mathit}{cmr}{bx}{it}
  \newmathalphabet{\mathbfit} 
  \addtoversion{normal}{\mathbfit}{cmr}{bx}{it}
  \addtoversion{bold}{\mathbfit}{cmr}{bx}{it}
  \newmathalphabet{\mathbfss} 
  \addtoversion{normal}{\mathbfss}{cmss}{bx}{n}
  \addtoversion{bold}{\mathbfss}{cmss}{bx}{n}
  \ifAMStwofonts
    \ifCUPmtlplainloaded \else
      \UseAMStwoboldmath
      \makeatletter
      \new@mathgroup\upmath@group
      \define@mathgroup\mv@normal\upmath@group{eur}{m}{n}
      \define@mathgroup\mv@bold\upmath@group{eur}{b}{n}
      \edef\UPM{\hexnumber\upmath@group}
      \new@mathgroup\amsa@group
      \define@mathgroup\mv@normal\amsa@group{msa}{m}{n}
      \define@mathgroup\mv@bold\amsa@group{msa}{m}{n}
      \edef\AMSa{\hexnumber\amsa@group}
      \makeatother
      \mathchardef\upi="0\UPM19
      \mathchardef\umu="0\UPM16
      \mathchardef\upartial="0\UPM40
      \mathchardef\leqslant="3\AMSa36
      \mathchardef\geqslant="3\AMSa3E

       \let\le=\leqslant

    \fi
  \fi
\fi 

\ifnfsstwo
  \DeclareMathAlphabet{\mathbfit}{OT1}{cmr}{bx}{it}
  \SetMathAlphabet\mathbfit{bold}{OT1}{cmr}{bx}{it}
  \DeclareMathAlphabet{\mathbfss}{OT1}{cmss}{bx}{n}
  \SetMathAlphabet\mathbfss{bold}{OT1}{cmss}{bx}{n}
  \ifAMStwofonts
    \ifCUPmtlplainloaded \else
      \DeclareSymbolFont{UPM}{U}{eur}{m}{n}
      \SetSymbolFont{UPM}{bold}{U}{eur}{b}{n}
      \DeclareSymbolFont{AMSa}{U}{msa}{m}{n}
      \DeclareMathSymbol{\upi}{0}{UPM}{"19}
      \DeclareMathSymbol{\umu}{0}{UPM}{"16}
      \DeclareMathSymbol{\upartial}{0}{UPM}{"40}
      \DeclareMathSymbol{\leqslant}{3}{AMSa}{"36}
      \DeclareMathSymbol{\geqslant}{3}{AMSa}{"3E}

       \let\le=\leqslant

    \fi
  \fi
\fi 

\ifCUPmtlplainloaded \else
  \ifAMStwofonts \else 
    \def\upi{\pi}
    \def\umu{\mu}
    \def\upartial{\partial}
  \fi
\fi

\title{Substructures in Cold Dark Matter Haloes}
\author[G.~De Lucia et al.]
        {G.~De Lucia,$^1$\thanks{Email: gdelucia@mpa-garching.mpg.de} 
        G.~Kauffmann,$^1$
        V.~Springel,$^1$
        S.~D.~M.~White,$^1$
        \newauthor B.~Lanzoni,$^2$
        F.~Stoehr,$^1$
        G.~Tormen,$^3$
        N.~Yoshida$^4$
        \\      
        $^1$Max--Planck--Institut f\"ur Astrophysik,
        Karl--Schwarzschild--Str. 1,85748 Garching bei M\"unchen, Germany 
        \\
        $^2$INAF--Osservatorio Astronomico di Bologna, via Ranzani 1, 40127
        Bologna, Italy 
        \\
        $^3$Dipartimento di Astronomia, Universita di Padova, vicolo dell
        Osservatorio 5, 35122 Padova, Italy
        \\
        $^4$Harvard--Smithsonian Center for Astrophysics, 60 Garden
        Street, Cambridge, MA02138}

\pagerange{\pageref{firstpage}--\pageref{lastpage}}
\pubyear{2003}

\begin{document}

\maketitle

\label{firstpage}

\begin{abstract}
We analyse the properties of substructures within dark matter haloes 
(subhalos) 
using a set of high-resolution numerical simulations of the formation of 
structure in a $\Lambda$CDM Universe.  Our simulation set includes $11$ 
high-resolution simulations of massive clusters as well as a region of mean 
density, allowing us to study the spatial and mass distribution of 
substructures down to a mass resolution limit of 
$10^{9}\,h^{-1}{\rm M}_{\odot}$. We also investigate how the properties of 
substructures vary as a function of the mass of the `parent' halo in which 
they are located.  We find that the substructure mass function depends at
most weakly on the mass of the parent halo and is well described by a
power-law. The radial number density profiles of substructures are steeper in
low mass haloes than in high mass haloes. More massive substructures tend to
avoid the centres of haloes and are preferentially located in the external 
regions of their parent haloes. 
We also study the mass accretion and merging histories of substructures, which
we find to be largely independent of environment. 
We find that a significant fraction of the substructures residing in clusters
at the present day were accreted at redshifts $z < 1$. This implies that a
significant fraction of present-day `passive' cluster galaxies were
still outside the cluster progenitor and were more active at $z \sim 1$.
\end{abstract}


\section[]{Introduction}
\label{sec:intro}

The formation and evolution of structure in the Universe is a topic of
fundamental interest.  In the last decades, the Cold Dark Matter (CDM)
model has been extensively studied and has had considerable success in
reproducing observational results, both on galactic and on cluster
scales.  In fact, the CDM model with the `concordance' set of cosmological 
parameters ($\Lambda$CDM) has been so successful that it can now be considered
a standard paradigm for the formation of structure in the Universe.  According
to this model, the dominant force that drives structure formation is gravity,
and large systems like galaxy clusters are formed via hierarchical merging 
of smaller structures. 

Numerical simulations of gravitational clustering of dark matter are
an indispensable tool for investigating the non-linear growth of
structures in its full geometrical complexity.  Until recently, dissipationless
simulations suffered from the so-called \emph{overmerging} problem,
i.e.~substructures disrupt very quickly within dense environments \citep{katz}.
However, both analytic work \citep*{moorekatz} and high resolution simulations
\citep*{tormen,ghigna,klypin,ghigna2} have demonstrated that the cores
of dark matter haloes that fall into a cluster can actually survive as
self-gravitating objects orbiting in the smooth dark matter background
of the cluster, provided high enough force and mass resolution are
used.  Recent high-resolution simulations \citep{ghigna2,volker2} have also
shown that the abundance of these substructures is in agreement with
the observed abundance of galaxies in clusters, suggesting a natural
one-to-one identification of luminous cluster galaxies and dark matter
substructures.

Another interesting claim is that the shape of the substructure mass
function is independent of the mass of the parent halo \citep{moore2}.
It is not obvious that this should be the case, because in CDM
cosmologies, the initial conditions do depend on scale and galaxies form 
several billion years before clusters.
As \citet{moore2} pointed out, the logarithmic slope of the power spectrum 
asymptotically approaches $-3$ on small scales, so clumps of widely 
different (yet sufficiently small) mass tend to collapse simultaneously, 
and as a result the timescale between the collapse of the first substructures 
and their incorporation into larger haloes becomes shorter.  
One might therefore expect that substructures were more easily disrupted in 
low mass haloes, producing a substructure mass function which depends on the 
mass of the parent halo. 

Observationally, the predicted abundance of substructures in clusters
is one of the major successes of the CDM model \citep{volker2}, but on
galactic scales, it appears that simulations predict more substructures 
than are visible by almost two orders of magnitude
\citep*{kauff,moore,klypin,tasi}.  This is commonly referred to as the `dwarf
galaxy crisis' of CDM.  There have been suggestions that the solution
to this problem lies in processes such as heating by a photo-ionising
background that suppresses star formation in small haloes at early
times \citep*{efsta,bull,som,benson}.
Alternatively, it has been suggested that the nature of dark matter may be
different than assumed in the canonical $\Lambda$CDM model, for example by
being {\em warm} or {\em self-interacting}, both of which could selectively
eliminate small-scale structure.  However,
self-interactions appear to be relatively ineffective in reducing the
number of subhalos, unless the assumed cross-section is unreasonably large
\citep{colin}.

A less drastic resolution was suggested by \citet{felix} who noted that it
might be possible to identify the observed Galactic satellites with the few
most massive subhalos and that the rest contain no stars. Direct evidence for
the large population of dark satellites predicted by CDM models comes from the
anomalous flux ratios of multiply imaged quasars \citep{mao,chiba,dalal}.

So far, a detailed numerical analysis of substructures has only been
carried out in high resolution re-simulations of a few individual
haloes \citep{moore2,ghigna2,volker2}.  In this paper, we carry out a 
systematic analysis of substructures as a function of the mass of the 
parent halo and as a function of environment.  We study the mass functions 
of subhalos, their radial distributions and their merging and mass
accretion histories. These quantities are of fundamental interest for
galaxy formation, because dark matter haloes and substructures
represent the birth places of luminous galaxies. Their accretion and
merging histories regulate the rate at which baryons can cool,
determining in this way the rate at which stars form in galaxies as a
function of cosmic time.

The layout of the paper is as follows: in Section~\ref{sec:simulations} we 
describe the simulations that we use; in Section~\ref{sec:subfind} we give 
a brief description of the algorithm used to find substructures in haloes; in
Section~\ref{sec:subMF} we present the subhalo mass function, in
Section~\ref{sec:massstat} we analyse the mass distribution of the largest 
substructures; in Section~\ref{sec:substat} we study the radial distribution 
of substructures; in Section~\ref{sec:hist} we discuss the merging and mass 
accretion histories of substructures, both as a function of mass and as a 
function of environment. A summary and a discussion of the results obtained 
are presented in Section~\ref{sec:concl}.


\section[]{Cluster simulations}
\label{sec:simulations}

In this study, we use collisionless simulations of clusters generated
using the `zoom' technique \citep{tormen,katz}. First,
a cosmological simulation of a large region is used to select a
suitable target cluster.  The particles in the final cluster and its
surroundings are then traced back to their initial Lagrangian region and are
replaced by a larger number of lower mass particles.  These are
perturbed using the same fluctuation distribution as in the parent
simulation, but now extended to smaller scales to account for the
increase in resolution. This resampling of the initial conditions of
the Lagrangian region of the cluster thus allows a localised increase
in mass and force resolution.  Outside the \emph{high-resolution}
region, particles of variable mass, increasing with distance, are
used so that the computational effort is concentrated on the cluster of
interest, while still maintaining a faithful representation of the
large-scale density and velocity field of the parent simulation.

In this paper, we study a set of $11$ high-resolution re-simulations
of galaxy clusters ($5$ of mass $10^{14}\,h^{-1}{\rm M}_{\odot}$ and
$6$ of mass $10^{15}\,h^{-1}{\rm M}_{\odot}$), and a high resolution
re-simulation of a `typical' region of the Universe.  The simulations 
were carried out with the parallel tree-code {\small GADGET}
\citep*{volker1}.

One of our massive clusters was taken from the `S-series' studied by
\citet{volker2}, where the parent simulation employed was the
GIF-$\Lambda$CDM model carried out by the Virgo Consortium
\citep{k99}.  This parent simulation followed $256^3$ particles of
mass $1.4\times10^{10}\,h^{-1}{\rm M}_{\odot}$ within a comoving box of
size $141.3\, h^{-1}$Mpc on a side.  The other cluster re-simulations
and the simulation of the field region were selected from the
VLS simulation carried out by the Virgo Consortium \citep*{jenk,yoshida}.  
The simulation was performed using a parallel P3M code \citep{mf} and 
followed $512^3$ particles with a particle mass of 
$7\times 10^{10}\,h^{-1}\,{\rm M}_{\odot}$ in a comoving box of size 
$479\,h^{-1}$Mpc on a side.   In all cases, the parent simulation and
the re-simulations were characterised by the following cosmological
parameters: $\Omega_0=0.3$, $\Omega_{\Lambda}=0.7$, spectral shape
$\Gamma=0.21$, $h=0.7$ (we adopt the convention $H_0=100\,h\, {\rm km
\, s^{-1}Mpc^{-1}}$) and normalisation $\sigma_8=0.9$.

In Table \ref{tab:tab1}, we summarise some important numerical parameters of 
the simulations used. We will refer to our five high-mass clusters of mass 
resolution $2\times10^{9}\,h^{-1}{\rm M}_{\odot}$ as type `B1', and to the 
low-mass clusters as type `B2'. These simulations were carried out by 
Barbara Lanzoni as part of her PhD thesis and were previously used in
\citet{lanzoni}. The `S2' simulation is taken from the `S-series' of
\citet{volker2}. The field region `M3' was adopted from the `M-series' studied
by \citet{felixthesis}. 

\begin{table*}
\caption{Numerical parameters for the simulations used. All the
  simulations were carried out assuming a $\Lambda$CDM cosmology with
  cosmological parameters $\Omega_0=0.3$, $\Omega_{\Lambda}=0.7$,
  $\Gamma=0.21$, $\sigma_8=0.9$, and $h=0.7$. In the table, we give the
  particle mass $m_{\rm p}$ in the high resolution region, the
  starting redshift $z_{\rm start}$, the gravitational softening
  $\epsilon$ in the high-resolution region and the number of simulations in
  each group $N$.}

\begin{tabular}{llllll}
\hline
Name & Description & $m_{\rm p}$ [$h^{-1}$M$_{\odot}$] & $z_{\rm start}$ &
$\epsilon$ [$h^{-1}$kpc] & $N$ \\
\hline

B$1$ & $10^{15}\,h^{-1}{\rm M}_{\odot}$ clusters & $2 \times 10^9$ & 60 & 5.0  &
$5$\\

B$2$ & $10^{14}\,h^{-1}{\rm M}_{\odot}$ clusters & $2 \times 10^9$ & 60 & 5.0  &
$5$ \\

S$2$ & $10^{15}\,h^{-1}{\rm M}_{\odot}$ cluster & $1.36 \times 10^9$ & 50 & 3.0
& $1$ \\

M$3$ & field simulation & $1.7 \times 10^8$ & 120 & 1.4  & $1$\\
\hline
\end{tabular}
\label{tab:tab1}
\end{table*}


\section[]{Identification of dark matter substructures}
\label{sec:subfind}

The identification of substructures in dark matter haloes is a
difficult technical problem and many different algorithms have been
developed to accomplish this task, for example the hierarchical
friends-of-friends algorithm (HFOF) \citep*{gott}, the bound density
maximum algorithm (BDM) \citep{kly}, and SKID (see http ref:
http://www-hpcc.astro.washington.edu/tools).  Each of these algorithms
has its own advantages and weaknesses, so that arguably none of
them is completely satisfactory yet.  In this work, we use the
algorithm {\small SUBFIND} proposed by \citet{volker2}, which combines
ideas used in other group finding techniques with a topological
approach for finding substructure candidates.  {\small SUBFIND}
can handle haloes of arbitrary shape, does not require an iterative
procedure for finding subhalo candidates, and is capable of detecting
arbitrary levels of `subhalos within subhalos'. In this section, we
briefly summarise how the method works.

In a first step, a standard friends-of-friends (FOF) algorithm is used
to identify virialized parent haloes. The FOF algorithm links together
all particle pairs with separation less than a linking length $b$. We
adopt the standard value $b=0.2$ in units of the mean particle
separation, which selects groups of particles with overdensities
close to the value predicted by the spherical collapse model for the
virialized regions of haloes.  The next step is to compute an estimate
of the local density at the position of each particle in the group. To
this end, we employ an adaptive kernel interpolation method similar to
the one used in smoothed particle hydrodynamics.  In the resulting
density field, we define as {\em substructure candidates} locally
overdense regions which are enclosed by isodensity contours that
traverse a saddle point.  Our method for finding these regions can be
visualised as follows: we reconstruct the density field by considering
particles in order of decreasing density, thus working our way from
high to low density.  This corresponds to gradually lowering a global
threshold in the density field sampled by the simulation
particles. Isolated overdense regions grow slowly in size during this
process.  When two such separate regions coalesce to form a single
region, their density contours join at a saddle point. Each time such
an event occurs, we have found two substructure candidates.
 
After the regions containing substructure candidates have been identified, we 
apply an unbinding procedure where we iteratively reject all particles with
positive total energy in order to eliminate `background' particles that do not
belong to the subhalo. For the purposes of this study, we consider all
substructures that survive this unbinding procedure, and still have at least
$10$ self-bound particles, to be genuine subhalos. 

In summary, the algorithm {\small SUBFIND} decomposes a given particle group 
into a set of disjoint and self-bound substructures, each of which is 
identified as a locally overdense region in the density field of the 
background halo. Note that {\small SUBFIND} classifies all the particles 
inside a FOF group either as belonging to a bound substructure or as being 
unbound.  The self-bound part of the FOF background halo itself will then 
also appear in the substructure list.
We will exclude it when referring to subhalos or substructures in the  
following analysis.


\section[]{The subhalo mass function}
\label{sec:subMF}

The sample of parent haloes used for studying the mass function analysis
consists of $6$, $5$, $34$ and $100$ haloes in the mass ranges
$8.68\times10^{14}$--$1.79\times10^{15}$ (from simulations B1 and S2), 
$6.99\times10^{13}$--$1.27\times10^{14}$ (from simulations B2), 
$7.0\times10^{12}$--$2.0\times10^{13}$ (from simulation M3) and 
$7.0\times10^{11}$--$2.0\times10^{12}\,h^{-1}{\rm M}_{\odot}$ 
(from simulation M3).

\begin{figure*}
\centering
\epsfig{file=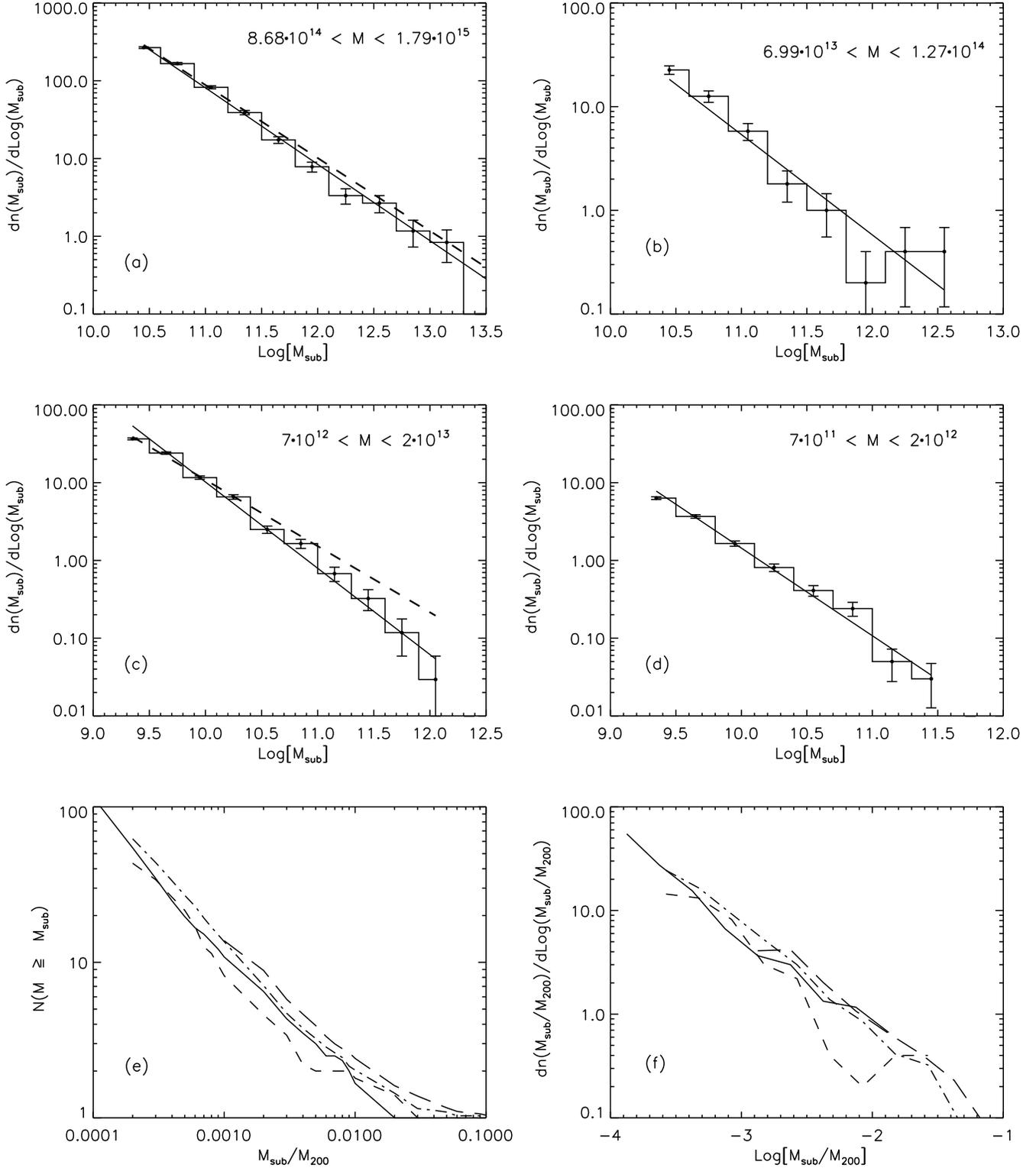}
\caption{Panel (a) -- (d): the differential mass function of subhalos 
  residing in parent haloes of different mass. The solid line represents a
  power law fit to the mass function. The dashed line shown in panel (a) and
  (c) 
  represents a fit restricted to the $4$ lowest mass bins.
  The masses are in units of $h^{-1}{\rm M}_{\odot}$. 
  The range of mass of the haloes used in each bin is indicated in each
  panel.
  Panel (e): the cumulative mass function of subhalos in units of rescaled 
  subhalo mass. The solid line is for haloes with mass 
  $\simeq 10^{15}\,h^{-1}{\rm M}_{\odot}$, the dashed line is for haloes with 
  mass $\simeq 10^{14}\,h^{-1}{\rm M}_{\odot}$, the dash-dotted line is for 
  haloes with mass $\simeq 10^{13}\,h^{-1}{\rm M}_{\odot}$, and the 
  long-dashed line is for haloes with mass $\simeq 10^{12}\,h^{-1}{\rm
  M}_{\odot}$.  
  Panel (f): differential mass functions in units of rescaled subhalo mass; 
  the different line styles are the same as in panel (e).}
\label{fig:fig1}
\end{figure*}

The resulting subhalo mass functions are shown in Fig.~\ref{fig:fig1}.
In the first four panels, we plot the differential mass functions
for parent haloes of different mass.  The histograms are computed by
stacking all the haloes in the given range of mass and the error bars 
represent Poisson errors. The solid line in each of the panels is a 
power-law fit to the measured differential mass function; the fit is 
performed using the least absolute deviation method over the range of 
mass shown by the line. In all the cases the slope of this unrestricted 
fit is close to $-1$ (it is equal to $-0.98$ for the top left panel, 
$-0.97$ for the top right panel, $-1.11$ for the middle left panel and 
$-1.13$ for the middle right panel). However, we note that the lowest mass
bins, which have the smallest statistical errors, are best fit with a
slightly shallower slope: if we restrict the fit to the $4$ lowest mass 
bins the slope is $-0.94$ for the top left panel and $-0.85$ for the
middle left panel. These are closer to the value $-0.8$, measured by 
\citet*{amina} for a single cluster simulation of extremely high-resolution. 
Also note that a slope shallower than $-1$ at the low-mass end implies that 
the integrated mass in substructures remains bounded and is 
dominated by the most massive subhalos. It is likely that our 
subhalo mass functions are steepened somewhat by a cut-off in 
abundance for very massive substructures.

The bottom left panel of Fig.~\ref{fig:fig1} shows the cumulative mass function
for all the haloes used in the sample.  To compare the different subhalo mass
functions, we have rescaled the subhalo mass by dividing by the virial
mass of the parent halo.  Each line represents the average cumulative mass
function over all the haloes in each mass bin. 
Note that in this paper we define the `virial mass' as $M_{200}$, the mass
within a sphere of density $200$ times the critical value at redshift zero.
The lines end at different places because of 
the differing mass resolution of the simulations (see Table
\ref{tab:tab1}). We find that all four cumulative mass functions agree
within the statistical errors.  Finally, in the bottom right panel, we
show the differential subhalo mass functions in units of rescaled
mass.

\begin{figure}
\centering
\epsfig{file=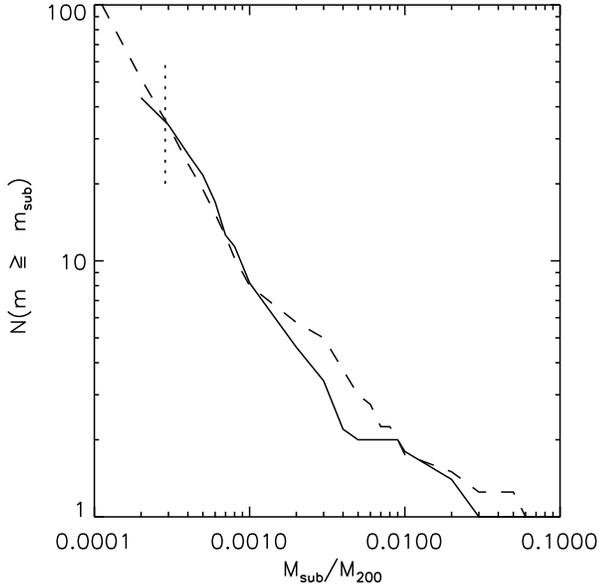}
\caption{The cumulative mass function of subhalos in units of rescaled
  subhalo mass for haloes with mass $\simeq 10^{14}\,h^{-1}{\rm
  M}_{\odot}$. The solid line corresponds to the average of haloes from
  simulations B2 and the dashed line to haloes from simulation M3.  The
  vertical dotted line shows the resolution limit corresponding to
  simulations B2.}
\label{fig:fig1b}
\end{figure}

We note that the `universality' of the subhalo mass function seen
here appears to be quite robust with respect to numerical
resolution. In Fig.~\ref{fig:fig1b} we compare the average cumulative
mass functions for haloes with mass $\simeq 10^{14}\,h^{-1}{\rm
M}_{\odot}$ from simulations B2 and M3. We here averaged $5$ haloes for
simulations B2, and $4$ for simulation M3, to reduce the
object-to-object scatter that is unavoidable for small numbers of subhalos.
Despite an order of magnitude difference in numerical
resolution, the agreement between the simulations is good. 
We are able to resolve the right number of objects
in the low-resolution simulation above its resolution limit (shown as
a vertical dotted line in the figure). A similar result was obtained
by \citet[][see their Fig.~5]{volker2} using a set of $4$
re--simulations of the same cluster with systematically increasing
resolution, thereby allowing a direct study of numerical convergence.
This showed in particular that the S2 simulation used here has well
converged to the properties of a much higher resolution simulation 
above its own resolution limit, as used here.
Further support for our results was also found by \citet[][see their
Fig.~3]{felix2}. They compared the S--series simulations from
\citet{volker2} with an extremely well resolved re--simulation of a
Milky--Way sized halo. This latter simulation used an updated version
of the simulation code and more conservative integration parameters
than used here \citep[following][]{power}, suggesting that the subhalo
mass function is a relatively robust quantity and that the simulations
we discuss here are adequate for our purposes.
  
\begin{figure}
\centering
\epsfig{file=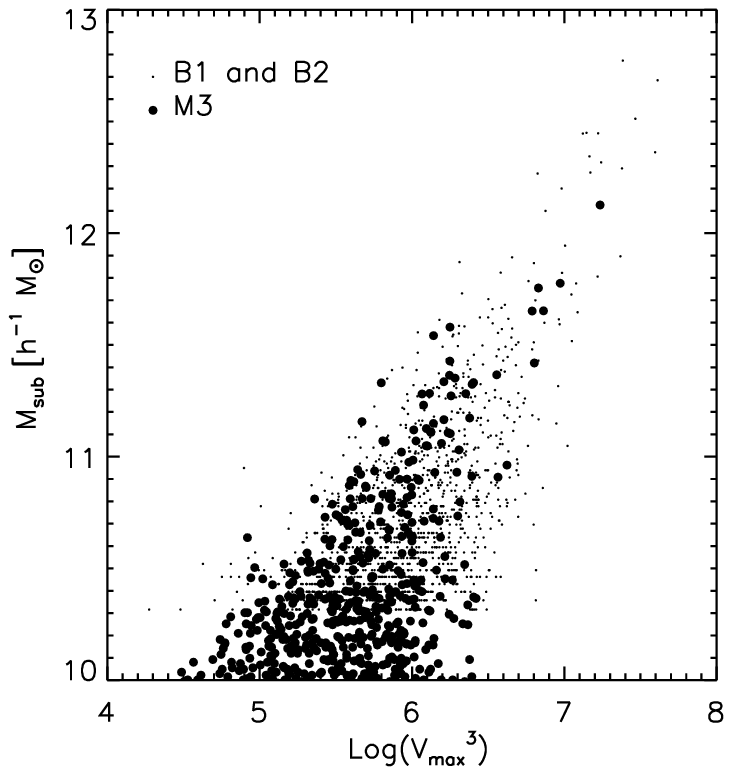}
\caption{Substructure masses as a function of $V_{\rm max}^3$ for
  subhalos drawn from our different simulations.  Small dots are used for
  subhalos that reside in haloes with mass $\sim10^{15}\,h^{-1}\,M_{\odot}$
  (from simulations B1) and $\sim10^{14}\,h^{-1}\,M_{\odot}$ (simulations B2)
  and filled circles for subhalos in haloes with mass
  $\sim10^{13}\,h^{-1}\,M_{\odot}$ (simulation M3). $V_{\rm max}$ was
  determined as the maximum of the circular velocity curve of each subhalo.}
\label{fig:fig1c}
\end{figure}

As a further check of the robustness of our results we also compare the
internal structure of subhalos drawn from our different simulations. 
Fig.~\ref{fig:fig1c} shows the correlation between the substructure mass and
the third power of the maximum circular velocity, $V_{\rm max}$, measured
directly from the circular velocity curve of each subhalo.  Different symbols
are used for substructures drawn from different simulations.  Note that for the
range of masses shown in the plot, substructures drawn from simulation M3
contain at least $60$ particles. While the scatter is clearly large for haloes
with such a low number of particles, the good general agreement between 
the runs suggests that the smallest substructures in our lower resolution
simulations have an internal structure that is still reliably resolved, at
least in a statistical fashion. 

Our results confirm the conclusion drawn by \citet{moore2}: the mass
function of substructures appears to be almost independent of the mass
of the parent halo. While our results are consistent with such a
`scale--free' subhalo mass function, the halo--to--halo scatter in our
simulation set is quite large, preventing us from putting tight
constraints on the accuracy with which the `scale--free' subhalo mass
function is preserved when haloes of different mass are
considered. And hence there is still room for weak trends with mass. A
clear detection of these would require simulations with larger
dynamic range, and larger samples of simulated haloes for each
mass bin.

As we discuss in more detail in Sec.~\ref{sec:mah}, our findings
suggest that the destruction of satellites due to the physical
processes of dynamical friction and tidal stripping on one hand, and
the accretion of new satellites on the other hand, cancel out in such
a way that the subhalo mass function does not depend or at best very weakly
depends on the mass of the parent halo. The reason for the invariance
of the subhalo mass function may lie in the physical nature of
this dynamical balance, which may be insensitive to the slightly
broken scale-invariance of dark haloes themselves. This shows up as a
mass-dependence of halo concentrations, for example.  Some fully analytic
models 
for the subhalo abundance have been developed \citep[e.g.][]{ravi}, but
they are presently not able to account for mass-loss and
dynamical friction self-consistently, and so provide little guidance to
answer this interesting question. A full understanding of the apparent
`conspiracy' that establishes an almost mass-invariant subhalo mass function
will therefore require further simulations.

\section[]{The most massive substructures}
\label{sec:massstat}

In this section we investigate whether the properties of the largest
substructures depend on the mass of the parent halo.  This is interesting 
since the largest substructures mark the sites where one expects to find the
brightest galaxies.   

In the following, $M_1$ refers to the mass of the most massive subhalo
and $M_2$ to the mass of the second most massive subhalo within
the virial radius of a given object of virial mass $M_{200}$. Note that
we have excluded from our analysis the subhalo associated with the FOF 
group itself. In a semi-analytic scheme, it is this `subhalo' that would host
the brightest cluster galaxy (BCG). In Sec.~\ref{sec:mah} we will show that,
once accreted onto a massive halo, substructures suffer significant stripping,
an effect that is more important for substructures accreted at higher redshift.
It is then likely that the largest substructures we find within the virial
radius at the present time were accreted at relatively low redshift. 

\begin{figure}
\centering
\epsfig{file=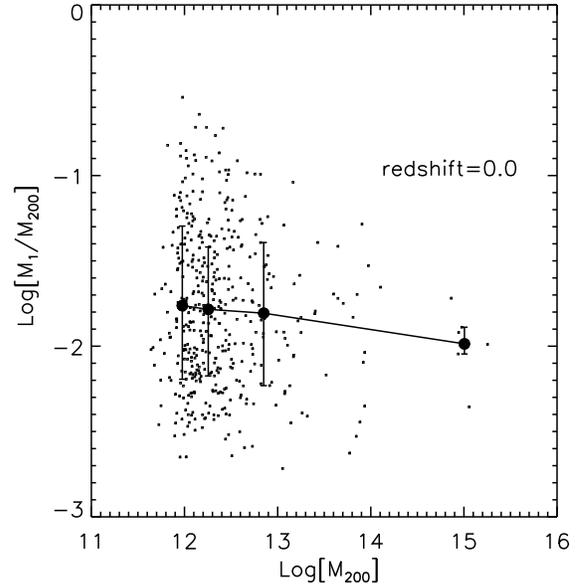}
\caption{Ratio of the mass of the most massive substructure and the
  parent halo mass as a function of parent halo mass.  The small symbols
  represent the values measured for each individual halo; the filled
  circles are the median in bins of halo mass chosen so that each of
  them contains the same number of points (143). The last six points,
  corresponding to the main haloes of simulations B1 and S2, are treated 
  as a separate bin. The error bars mark the $20$ and $80$ per cent 
  percentiles of the distribution.}
\label{fig:fig2}
\end{figure}

In Fig.~\ref{fig:fig2}, we plot $M_1/M_{200}$ as a function of $M_{200}$
for $434$ haloes drawn from all the simulations listed in Table~\ref{tab:tab1}.
This sample includes not only the central clusters in our re-simulations, 
but also the other haloes found in the high-resolution regions around the 
re-simulated objects down to a mass limit of $10^{13}\,h^{-1}\,{\rm
  M}_{\odot}$.  
We took care however to exclude \emph{contaminated} 
haloes that contained low resolution particles. In simulation M$3$, we 
selected only haloes with a mass larger than 
$10^{12}\,h^{-1}{\rm M}_{\odot}$.   
The small symbols in Fig.~\ref{fig:fig2} indicate the value of 
$M_1/M_{200}$ measured for each individual halo, while the filled circles 
represent the median of the distribution. We have taken bins in $M_{200}$ 
such that there are an equal number ($143$) of haloes in each bin, except 
for the last six points, which we treated as a separate bin, corresponding 
to the central cluster haloes in simulations B$1$ and S$2$.  
The error bars mark the $20$ and $80$ per cent percentiles of the distribution.

The results in Fig.~\ref{fig:fig2} suggest that $M_1/M_{200}$ depends very 
little on the mass of the parent halo. Interestingly, $M_1/M_{200}$ appears 
to exhibit less scatter for the most massive haloes, but the number of 
simulated clusters we have in this high mass regime is rather small, so it 
is unclear whether this effect is statistically significant.

The results in Fig.~\ref{fig:fig2} imply that the median value of 
$M_1$ increases in proportion to the mass of the cluster, suggesting that 
second ranked galaxies will be more luminous in more massive haloes. 
Note also that 
the mass of the largest substructure within the virial radius is typically 
only a few per cent of the virial mass.

\begin{figure}
\centering
\epsfig{file=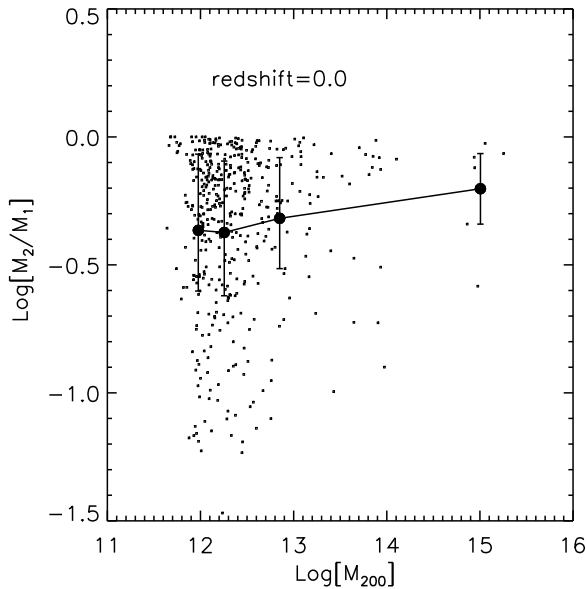}
\caption{Ratio in mass between the two most massive substructures as a
  function of the parent halo mass. As in Fig.~\ref{fig:fig2}, small
  symbols represent the values measured for each individual halo,
  while the filled circles give the median in the same bins as in
  Fig.~\ref{fig:fig2}. 
  The error bars mark the $20$ and $80$ per cent percentiles of the
  distribution.} 
\label{fig:fig3}
\end{figure}

In Fig.~\ref{fig:fig3}, we show the ratio $M_2/M_1$ as a function of the mass 
of the parent halo. Once again there is rather little dependence on 
$M_{200}$ with a possible decrease in the scatter for more massive haloes.  
Note that the median value of $M_2/M_1$ is around $0.5$. If the 
stellar masses of the second and third brightest galaxies in a cluster scale
simply with the masses of their dark subhalos, they should have K-band
luminosities that are equal to within $0.5$ mag.


\section[]{The radial distribution of subhalos}
\label{sec:substat}

\begin{figure}
\centering
\epsfig{file=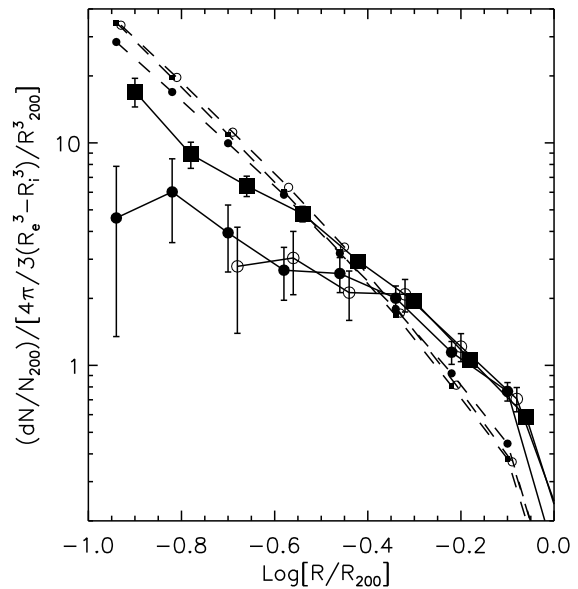}
\caption{Radial distribution of substructures in haloes of different
  mass. Lengths are given in units of $R_{200}$ and densities are
  normalised to the mean density inside $R_{200}$.  Symbols connected
  by solid lines show the number density profile of substructures
  (filled circles are for $10^{15}\,h^{-1}{\rm M}_{\odot}$ haloes, empty
  circles for $10^{14}\,h^{-1}{\rm M}_{\odot}$ haloes and filled squares for
  $10^{13}\,h^{-1}{\rm M}_{\odot}$ haloes). Symbols connected by dashed
  lines show the corresponding dark matter profiles.}
\label{fig:fig4}
\end{figure}

The large sample of subhalos in our simulations allows us to study
their radial distribution and to investigate how it depends on the
mass of the parent halo.  In Fig.~\ref{fig:fig4}, we plot the number
density of substructures as a function of the normalised distance
$R/R_{200}$ from the centre of the halo, defined here as the position
of the most bound particle in the halo.  We show averaged results for
haloes with masses $\sim 10^{15}\,h^{-1}{\rm M}_{\odot}$, $\sim
10^{14}\,h^{-1}{\rm M}_{\odot}$, and $\sim 10^{13}\,h^{-1}{\rm
M}_{\odot}$, and we limit the analysis to subhalos with masses greater
than a fixed fraction ($2\times 10^{-4}$) of the virial mass of the
parent halo. This fraction is chosen because it lies just above the
mass limit where it is possible to identify substructures in all of
our simulations.  As Fig.~\ref{fig:fig1} shows, there are typically
$\sim 50$ subhalos per parent halo with $M_{\rm sub}/M_{200} > 2
\times 10^{-4}$, so by stacking a large sample of haloes, it is
possible to calculate an average density profile that has rather
little noise.  Note that the densities plotted in Fig.~\ref{fig:fig4}
have been normalised to the mean density inside the virial radius.
The solid lines with symbols show results for the $3$ different parent
halo mass ranges defined above.  For comparison, we have also plotted
the dark matter radial profiles as dashed lines.  Note that a small
shift in the abscissa has been added to make the plot more readable.

We find that the subhalo profiles are `anti-biased' relative to the
dark matter in the inner regions of the haloes. This agrees with the
results of \citet{ghigna2}.  
Surprisingly we also find that the radial number density 
profiles are steeper in low mass haloes than in high mass haloes,
a finding that reserves further investigation.

\begin{figure}
\centering
\epsfig{file=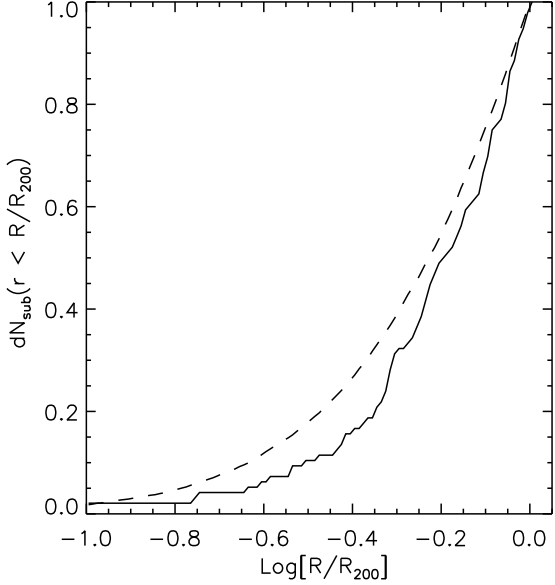}
\caption{Cumulative radial distribution of the number of substructures
  in two different mass ranges in the simulation M$3$. 
  The dashed line is for substructures with 
  $M_{\rm sub} < 0.01\,M_{200}$, and the solid line is for 
  $M_{\rm sub} >0.01\,M_{200}$.}
\label{fig:fig5}
\end{figure}

We now use our highest resolution cluster simulation to investigate
whether subhalos of different mass have different radial profiles.  In
Fig.~\ref{fig:fig5}, we show the cumulative fraction of substructures
as a function of $R/R_{200}$ for subhalos with $M_{\rm sub}>0.01\,M_{200}$ 
(solid line) and $M_{\rm sub}\le 0.01\,M_{200}$ (dashed line). 

As Fig.~\ref{fig:fig1} already made clear, there are many more
substructures with $M_{\rm sub}\le 0.01\,M_{200}$ than with $M_{\rm
sub}>0.01\,M_{200}$ ($9749$ versus $96$).  Fig.~\ref{fig:fig5} now
shows that more massive substructures are preferentially located in
the external regions of their parent haloes.  This can be understood as
a consequence of tidal truncation and stripping effects that quickly
decrease the mass of subhalos as they fall into the cluster and reach
the dense inner cores of the parent haloes (see Section~\ref{sec:mah}
for a more quantitative analysis of mass-loss due to stripping).
 
Also note that this finding can be naturally explained as a
consequence of the orbital decay experienced by substructures.  As
shown in \citet*{tordia}, the orbital decay is consistent with
expectations based on the combined effects of dynamical friction and
mass-loss.  As a result, massive substructures are driven to the
centre more rapidly than less massive ones: Tormen et al.~show that
the orbital decay occurs in less than a Hubble time if the initial
mass of substructures is larger than $1$ per cent of the mass of the
main cluster, while the substructures can retain their identity for a
significant fraction of the Hubble time if their mass is smaller than
$5$ per cent of the main cluster mass.  Once driven to the centre,
massive substructures are destroyed and no longer distinguishable from
the central halo; this naturally explains the mass segregation that we
see in our simulations.

\begin{figure}
\centering
\epsfig{file=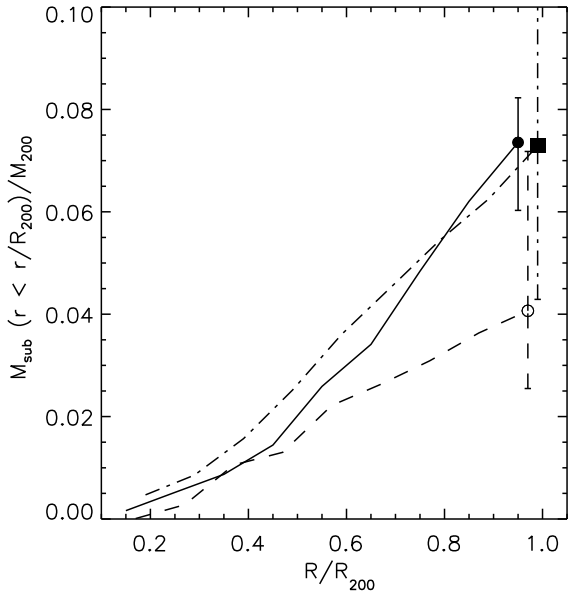}
\caption{Cumulative mass fraction in substructures as a function of the
  distance from the halo centre for haloes of different mass. The solid
  line is for haloes of mass $\simeq 10^{15}\,h^{-1}{\rm M}_{\odot}$, the
  dashed line for haloes of mass $\simeq 10^{14}\,h^{-1}{\rm M}_{\odot}$,
  and the dash-dotted line for haloes of mass 
  $\simeq 10^{13}\,h^{-1}{\rm M}_{\odot}$. The error bars on the last point
  mark the $20$ and $80$ percentiles of the distribution.}
\label{fig:fig6}
\end{figure}

In Fig.~\ref{fig:fig6}, we show the cumulative fraction of the total mass 
of the parent halo that is in substructures as a function of normalised 
distance from the halo centre. The different lines and symbols have the same 
meaning as in Fig. \ref{fig:fig4} and represent median relations for all the 
haloes in each mass bin. The error bars mark the $20$ and $80$ per cent 
percentiles of the distribution.  The mass fraction in substructures rises 
from $\simeq 1$ per cent at a radius $\simeq 0.3\,R_{200}$ to $\sim 6$ per 
cent at $r \sim R_{200}$. 
Note that the total mass fraction is dominated by the small number of most 
massive subhalos, and is hence a rather noisy quantity that shows large 
variations from system to system. The mass fraction may also reach values 
above $\simeq 10$--$15$ per cent, but then the underlying FOF parent halo is 
typically quite aspherical, with the most massive subhalo often lying 
outside the formal radius $R_{200}$ of the parent halo.

\section[]{Subhalo histories}
\label{sec:hist}

So far we have analysed the properties of subhalos only at redshift $z=0$. 
In this section, we turn to the time evolution of the masses of subhalos and 
their merging histories.

In order to carry out this analysis, we have measured \emph{merger trees} for 
each subhalo from the simulations. These trees allow us to specify all the 
progenitors (or the descendants) of a substructure at each epoch.  
To build the merger trees, we use a slightly modified version of the code 
described in \citet{volker2}. We briefly review the main features of the 
relevant algorithms in the following section.


\subsection{Constructing merging trees}
\label{sec:mergtrees}

Following \citet{volker2} we define a subhalo $S_{\rm B}$ at redshift
$z_{\rm B}$ to be the \emph{progenitor} of a subhalo $S_{\rm A}$ at
redshift $z_{\rm A}$ (with $z_{\rm A} < z_{\rm B}$) if more than one
half of the $N_{\rm link}$ most bound particles belonging to $S_{\rm
B}$ end up in $S_{\rm A}$. \citet{volker2} adopted $N_{\rm link}=10$.
However, we obtained considerably better results with a value of
$N_{\rm link}$ that varies between $10$ for the less massive
substructures to $100$ for more massive ones. In this way, occasional
failures of the code when building the merger trees were more robustly
avoided.  Particularly if substructures undergo major mergers, the
code identified in some cases the wrong progenitor for $N_{\rm
link}=10$, or lost track of an entire subhalo.  We have also
updated the code so that volatile links to `evanescent' substructures
(i.e.~objects close to the resolution limit that occasionally appear
and then disappear) are avoided.

With these choices, we manage to trace $85$--$90$ per cent of all the
substructures in our simulations back to the point when they were
first accreted.  This fraction goes up to $87$--$93$ per cent if we only
consider substructures with more than $100$ particles.


\subsection{Mass accretion history}
\label{sec:mah}

We now use our merging trees to study the \emph{mass accretion histories} 
of the subhalos in our simulations. \citet{frank} has carried out a similar 
analysis for dark matter haloes and has proposed an analytic expression 
for the  
mass accretion function based on the extended Press-Schechter formalism 
\citep{ps,bond,bower}. This function was found to be in excellent agreement 
with the results of high-resolution $N$-body simulations.  
Our aim is to study the mass accretion function for subhalos and study whether
there is any dependence on mass or on environment.

We have selected subhalos at redshift $z=0$ in two different mass ranges
($\simeq 10^{11}\,h^{-1}{\rm M}_{\odot}$ and $\simeq 10^{12}\,h^{-1}{\rm
  M}_{\odot}$). In order to test for the effects of environment, we selected on
one hand subhalos that reside within the virial radius of the massive clusters
that formed in simulations B$1$ and B$2$ by the present day, and on the other
hand subhalos located within the smaller haloes found in simulation M$3$. In
the following, we will refer to the substructures in the clusters as
\emph{cluster 
  subhalos} and to the substructures inside the smaller haloes of $M$3 as 
\emph{field subhalos}.  Note that since we have excluded from our analysis 
the main subhalo associated with the FOF group and since on average the most
massive substructure has a mass a few per cent of $M_{200}$ (see
Fig.~\ref{fig:fig2}) we end up with very few substructures selected from
M$3$. In particular we have only $5$ substructures with a mass $\sim
10^{12}\,h^{-1}{\rm M}_{\odot}$ and $38$ substructures with a mass $\sim
10^{11}\,h^{-1}{\rm M}_{\odot}$.  
The corresponding numbers for the substructures selected from simulations 
B$1$ and B$2$ are $62$ and $338$. 

For each of these samples, we build the mass accretion histories as
follows: we start from a particular subhalo at redshift $z=0$ and
construct its merger tree as described in the previous section. 
At each redshift we track the history of the selected subhalo by
linking it with its most massive progenitor.

\begin{figure}
\centering
\epsfig{file=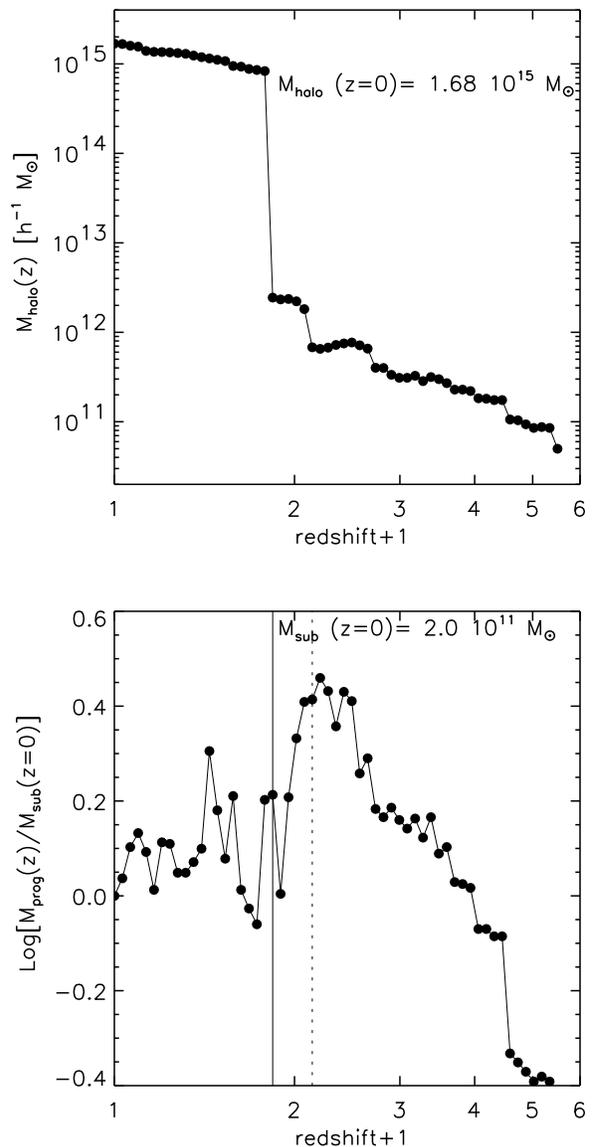}
\caption{Example for a typical mass accretion history for a subhalo of
  mass $2\times 10^{11} h^{-1}$M$_{\odot}$ (lower panel), and
  the corresponding variation of mass for the parent halo in which the
  subhalo resides (top panel). The vertical solid line corresponds to
  the last time the subhalo is outside the main progenitor of the
  cluster; the dotted line corresponds to the time the subhalo becomes
  a substructure (see the text for details).}
\label{fig:fig7}
\end{figure}

In Fig.~\ref{fig:fig7} we show a typical example of a mass accretion history
for a subhalo with mass  
$2\times 10^{11}\,h^{-1}{\rm M}_{\odot}$ at redshift zero.
In the lower panel, we show the mass accretion history of the subhalo
and in the upper panel, the corresponding mass of the halo in which the 
subhalo resides at each redshift.

In this example, the subhalo was accreted 
onto a larger halo at redshift $\sim 1$ (shown as a dotted line 
on the plot). For times prior to this event the substructure was a main
subhalo, i.e. the subhalo corresponding to a FOF group, and its mass 
grew monotonically in time. From now on, we will refer to this event
as the \emph{accretion time} ($t_{\rm accr}$) of the subhalo.  
A few snapshots later, at redshift $\sim 0.8$, the substructure and its host
halo were accreted onto the main progenitor of the cluster (shown as a solid
line on the plot).

After the subhalo is accreted, it suffers significant tidal stripping 
and {\em decreases} in mass. In this particular example, the final mass 
of the subhalo is $\simeq 40$ per cent of the value at its accretion time.

We find that for $\sim 60$ per cent of the subhalos in the $10^{11}$ mass bin
and $\sim 80$ per cent of the subhalos in the $10^{12}$ bin the accretion
event corresponds to the accretion of the substructure onto the main 
progenitor of the cluster itself. For most of the rest,
the time elapsed between these two events is fairly short. 
The results we will show later are essentially unchanged if we 
adopt as definition of the accretion time, the accretion of the substructure
onto the main progenitor of the cluster itself.

\begin{figure}
\centering
\epsfig{file=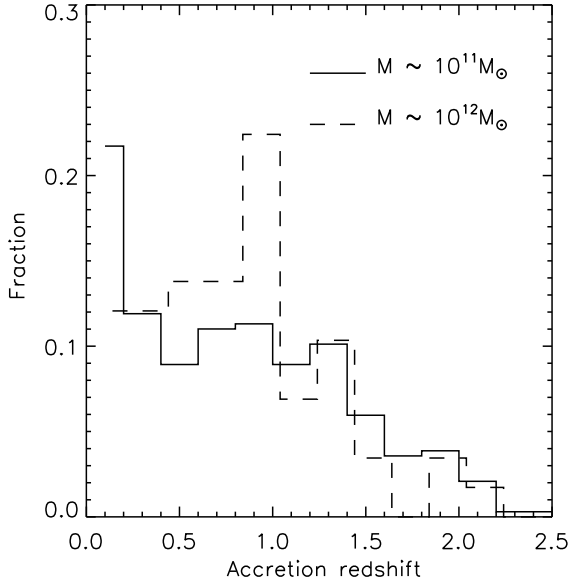}
\caption{Distribution of the accretion redshifts for the cluster
  subhalos sample. A small shift is added to the abscissa to produce a
  more readable plot.}
\label{fig:fig8}
\end{figure}

Fig.~\ref{fig:fig8} shows the distribution of the accretion redshifts for
the cluster substructures in our sample. Interestingly, we find that a 
large fraction of the substructures are accreted at redshift $z<1$. 
As noted above, for most of these substructures this accretion event
corresponds to the accretion onto the main cluster itself. Our results hint 
that substructures are constantly erased in the cluster, being replenished by
newly infalling haloes. 

\begin{figure}
\centering
\epsfig{file=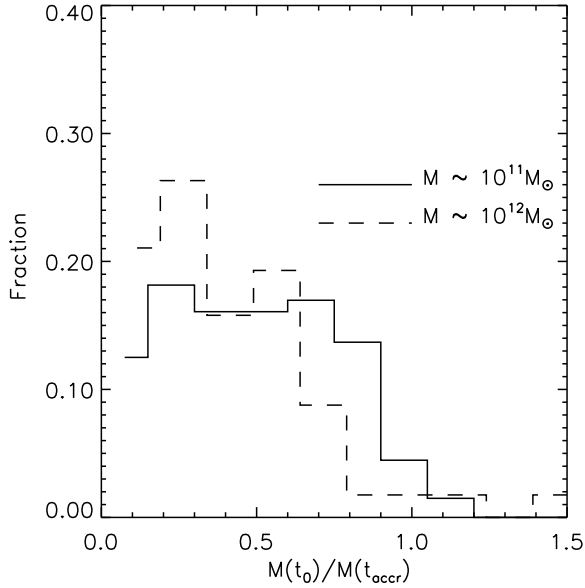}
\caption{Distribution of the quantity $M(t=t_0)/M(t=t_{\rm accr})$ for
  the cluster subhalo sample.  A small shift is added to the abscissa
  to produce a more readable plot.}
\label{fig:fig9}
\end{figure}

In Fig.~\ref{fig:fig9}, we plot the distribution of 
$M(t=t_0)/M(t=t_{\rm accr})$, i.e.~the ratio of the mass of the subhalo at 
the present day to the mass it had when it was first accreted. 
The histograms show that this ratio has a quite broad distribution, varying 
from a value of $\sim 1$ to $\sim 0.1$. We note that most of the subhalos 
that have lost only small amounts of mass have been accreted very 
recently. 

\begin{figure}
\centering
\epsfig{file=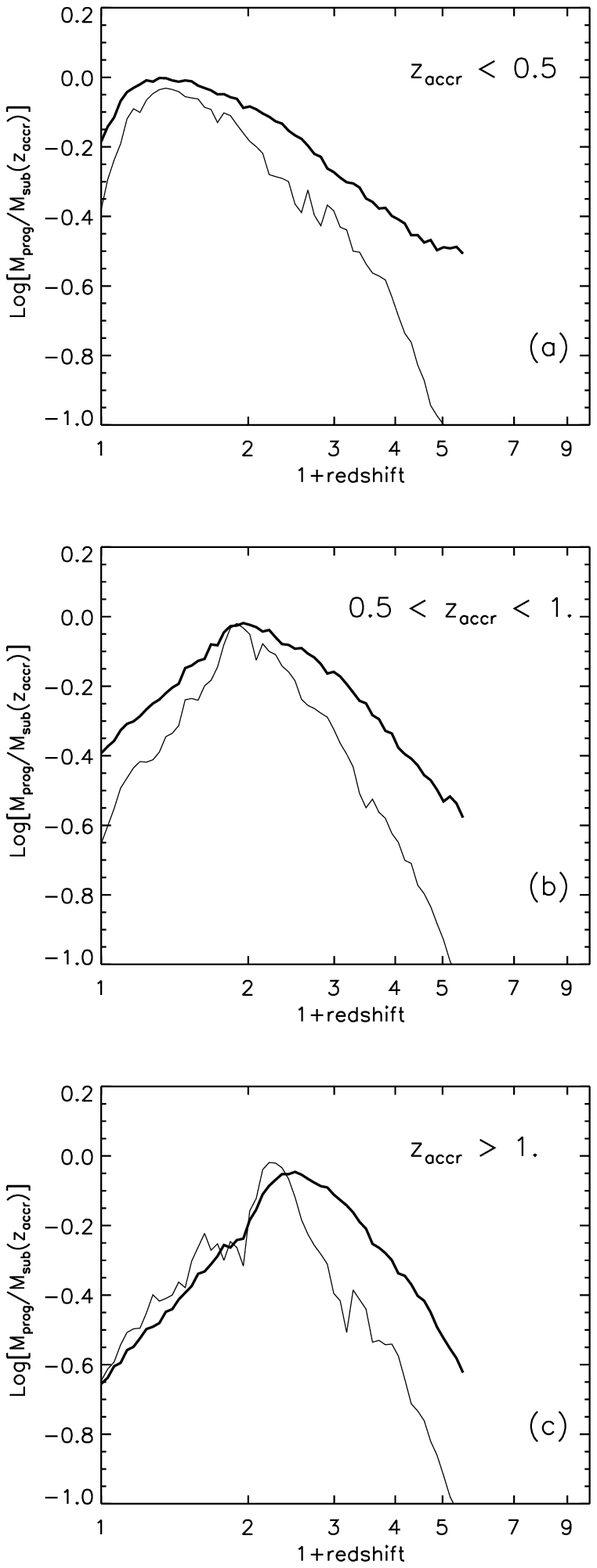}
\caption{Average mass accretion history for $z=0$ cluster subhalos accreted in
  three different redshift bins.
  Thick lines are used for subhalos with mass $\sim 10^{11}\,h^{-1}{\rm
    M}_{\odot}$ and thin lines are used for subhalos with mass $\sim
  10^{12}\,h^{-1}{\rm M}_{\odot}$. 
  The histories are normalised to the mass of the subhalo at the accretion
  time.}
\label{fig:fig10}
\end{figure}

This is more clearly shown in Fig.~\ref{fig:fig10} where we plot the average 
mass accretion function for the cluster subhalos in the two mass bins 
considered. Three different accretion redshift intervals are considered and 
in all cases the subhalo masses are normalised to the mass of the subhalo 
at $t_{\rm accr}$. 
The thick solid line shows the mean relation for subhalos with mass 
$\sim 10^{11}\,h^{-1}{\rm M}_{\odot}$, while the thin line shows the relation 
for $M_{\rm sub} \sim 10^{12}\,h^{-1}{\rm M}_{\odot}$. 
The mass accretion function
monotonically increases for times prior to the accretion event. Once the 
substructures are accreted, tidal stripping is effective and operates 
on short time scales. The longer the substructure spends in a more 
massive halo, the larger is the destructive effect of tidal stripping. 
Substructures remaining at $z=0$ that were accreted at redshift larger than $1$
(panel c) have been typically stripped of $\sim 80$ per cent of their
mass. There is also a slight indication that stripping is more effective for
more massive substructures: panels (a) and (b) show that more massive
substructures have been stripped significantly more than less massive
substructures accreted at the same redshift. 
This effect does not appear in panel (c) but note that we have very few 
objects accreted in this redshift bin in our sample of more massive subhalos. 

\begin{figure}
\centering
\epsfig{file=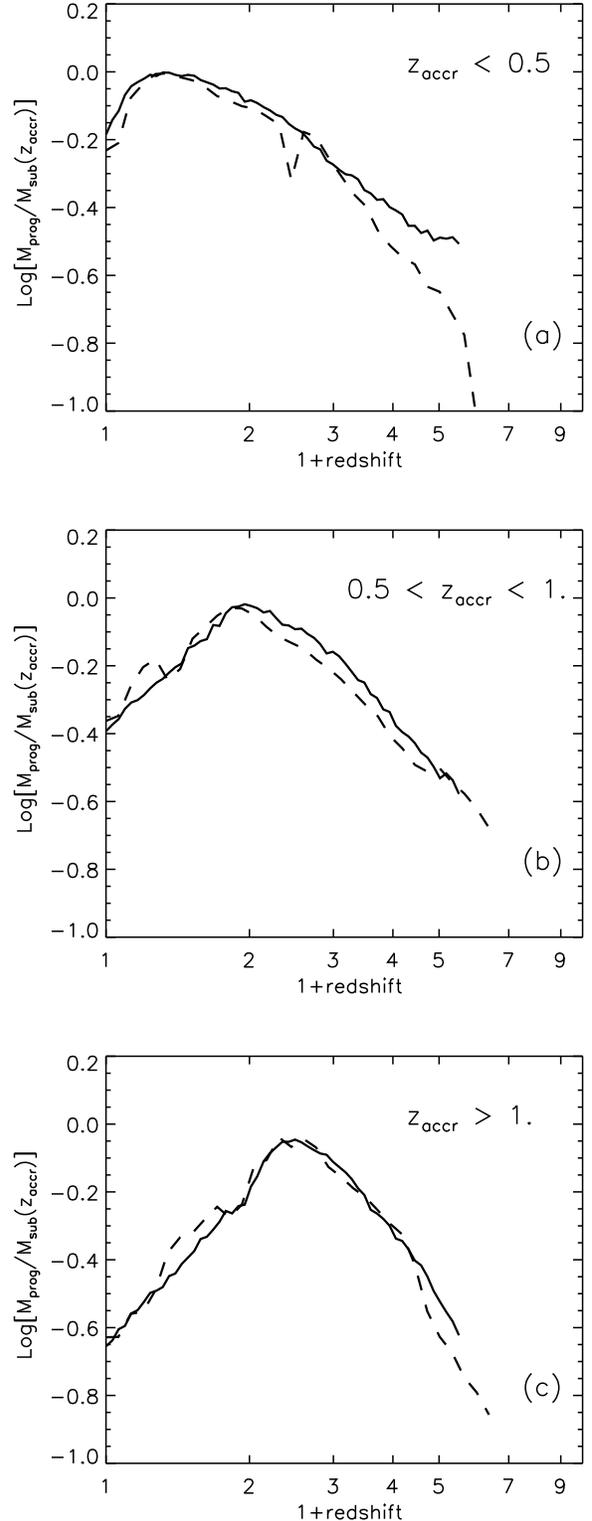}
\caption{Mass accretion histories for our lowest mass bin ($\sim
  10^{11}\,h^{-1}{\rm M}_{\odot}$) for cluster subhalos (solid line) and
  for field subhalos (dashed line). As in Fig.~\ref{fig:fig10}, 
  the sample is divided into three different subsamples according to the
  redshift of accretion.}
\label{fig:fig11}
\end{figure}

In Fig.~\ref{fig:fig11} we compare the mass accretion histories of field and 
cluster subhalos. We limit the analysis to substructures with mass 
$\sim 10^{11}\,h^{-1}{\rm M}_{\odot}$. Again the mass accretion function is
normalised to the mass of the subhalo at the accretion time. We find that field
subhalos and cluster subhalos have remarkably similar histories suggesting that
the efficiency of the tidal stripping is largely independent of the mass of the
parent halo.

\subsection{Merging histories}
\label{sec:mh}

In hierarchical models of galaxy formation, galaxies and their associated 
dark matter haloes form hierarchically through merger and accretion processes. 
In this context, the term \emph{merger} is usually used to refer to an 
interaction between two objects of similar mass, while the term 
\emph{accretion} is used to describe the infall of small objects 
onto much more massive haloes.

Observational results and numerical simulations confirm that interactions 
(such as tidal truncation or collisions) play an important role in the 
evolution of galaxies. There is, for example, solid evidence 
\citep{schweizer,whitmore,barnes} that at least some elliptical galaxies 
are the result of mergers between disk galaxies of similar mass.  
Mergers may also have a strong effect on the baryonic component of galaxies; 
they can trigger bar-instabilities in stellar disks and cause an inflow of 
gas into the galaxies centres, fuelling AGN activity or starbursts.

Following the merger tree of the substructures in our sample we can analyse in
detail the merger history, distinguishing between mergers and accretion events.
To build the merger history we proceed as follows: we consider all the 
substructures within the virial radius at redshift zero and follow their 
merger trees back in time, checking as substructures are accreted onto the 
main progenitor. 
We count as mergers all accretion events involving haloes with mass larger
than $2\times 10^{10}\,h^{-1}\,{\rm M}_{\odot}$ and mass ratio smaller
than $5\,:\,1$. Note that the lower limit on the mass of the merging haloes
correspond to the resolution limit of our cluster simulations 
(see Table \ref{tab:tab1}).

\begin{figure}
\centering
\epsfig{file=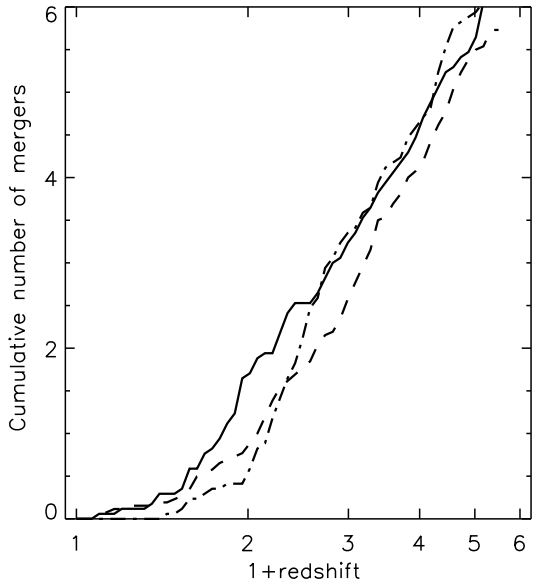}
\caption{The mean number of mergers after redshift $z$
  for substructures selected from the simulations B1 and B2 at redshift $z=0$
  and with mass $\sim 10^{12}\,h^{-1}{\rm M}_{\odot}$. The solid line is for
  substructures accreted at redshift $z < 0.5$, the dashed line for
  substructures accreted at redshift $0.5 < z < 1.0$ and the dashed-dotted line
  for substructures accreted at redshift $z > 1.0$.}
  \label{fig:fig12}
\end{figure}

In order to have enough information without running into numerical resolution
effects, we limit the present analysis to subhalos with mass $\sim
10^{12}\,h^{-1}\,{\rm M}_{\odot}$. 
Fig.~\ref{fig:fig12} shows the mean number of mergers per subhalo,
identified at redshift $z=0$, occurring after the redshift plotted in the
abscissa. 
As in the previous section, the sample is split into three different accretion
redshift intervals.

Merger events are less frequent once a halo is accreted onto 
a more massive structure. This is because the merging efficiency is higher in 
environments where the relative velocities of subhalos are similar to 
their internal virial velocities.  Once haloes are accreted by a massive halo, 
merging is suppressed by the large velocity dispersion they aquire. 
This effect is clearly visible in Fig.~\ref{fig:fig12} in the change in
slope of the curves near the accretion redshift. Note however that when one
goes to significantly higher redshift the difference between the three curves
vanishes.

\begin{figure}
\centering
\epsfig{file=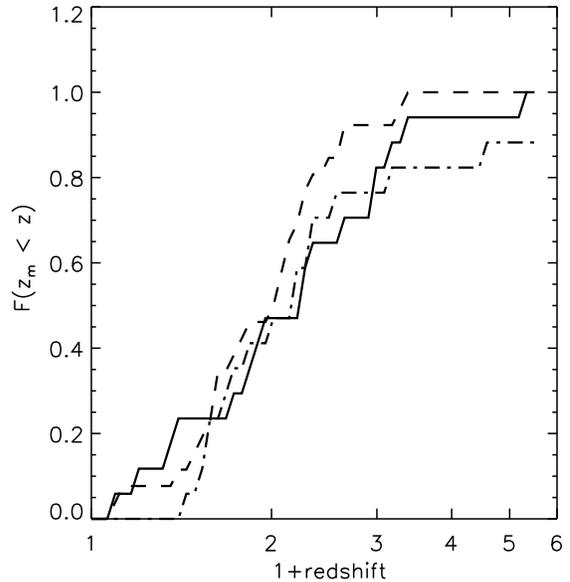}
\caption{Fraction of substructures with at least one merger event at redshift
  $z_m < z$. Different line styles are for different accretion redshift as in Fig.~\ref{fig:fig12}.}
\label{fig:fig13}
\end{figure}

In Fig.~\ref{fig:fig13} we show the fraction of subhalos that have had at least
one merger after the redshift plotted in the abscissa. Note that the merger 
events we are considering are characterised by similar masses and will most
likely influence the morphology of the main galaxy, leading to the formation of
a bulge component.  
The final morphology of the galaxies residing in these substructures will
depend on the time between the last major merger and accretion onto the
cluster: the longer this time, the larger is the likelihood that the galaxy
can grow a new disk.  

The results in Fig.~\ref{fig:fig13} show that $\sim 80$ per cent of the
$\sim 10^{12}\,h^{-1}\,{\rm M}_{\odot}$ haloes in our $z=0$ sample have had at
least one major merger at redshift below $2$; this fraction decreases to $\sim
40$ per cent for redshift $<1$. Surprisingly the fraction is almost independent
of the accretion time.  
These results suggest that a large fraction of subhalos in our cluster sample
will host early-type galaxies with a significant bulge component. 

\section{Summary and discussion}
\label{sec:concl}

We have used a set of high resolution numerical simulations in a $\Lambda$CDM 
Universe to study dark matter halo substructures.  Such dark matter
substructures mark the sites where luminous galaxies are expected to be found,
so the analysis of their mass functions, radial distributions, merging and mass
accretion histories should help us to better understand the
properties of the galaxies that form in hierarchical galaxy formation
models. Comparison with observational data should then suggest the physics that
needs to be included in viable models of galaxy formation and evolution.

In agreement with previous work \citep{moore2}, we find that the shape of the 
subhalo mass function is almost independent of the mass of the parent halo,
with galactic haloes being essentially scaled versions of cluster haloes.

We find that the average mass of the largest substructure within the
virial radius (excluding the BCG) scales linearly with the mass of the
parent halo.  If the stellar masses scale linearly with the dark
matter mass of the parent substructure, the second ranked galaxies
should have K-band luminosities that increase roughly linearly with
the mass of the main halo and are equal to those of the third ranked
galaxies to within $0.5$ mag.  

Note that the assumption that the stellar mass scales linearly with 
the dark matter mass of the parent substructure cannot generally be 
true since stars are typically much more concentrated than dark matter.  
The relation between stellar mass and the mass of parent substructure
is then quite complex and should be followed considering the details
of star formation and feedback process as is done for example in
\citet{volker2}.

We have also used the simulations to study the radial distribution of
substructures.  In agreement with previous work \citep{ghigna2}, we find that 
the radial profile of substructure number density is `anti-biased' relative 
to the dark matter profile in the inner regions of haloes. 
The most massive substructures reside preferentially in the outer regions of 
haloes. This is, at least in part, because substructures undergo substantial
tidal stripping in the dense inner regions of haloes.  

We have studied the evolution with time of this stripping process and find that
once a subhalo is accreted by a larger system, tidal stripping is highly
effective; the longer a substructure spends in a more massive halo, the larger
is the destructive effect. 
This suggests that substructures are constantly erased in clusters, being
replenished by newly infalling galaxies.  

Interestingly, a significant fraction of the substructures found in 
present-day clusters were first accreted at redshifts $z < 1$, 
implying that tidal truncation of the dark haloes of cluster 
galaxies happened relatively recently, and that at $z >1$ many 
present-day `passive' cluster members should still have been central galaxies
of their own  
extended dark haloes. If gas was able to cool in these haloes, the 
galaxies may have been considerably more active at $z\sim 1$ than at present.
Substructures in smaller haloes have histories remarkably similar
to those in cluster haloes, suggesting that the efficiency of tidal
stripping is largely independent of the mass of the main halo. 

Once a substructure is accreted onto a cluster, its merging probability 
decreases because of the large velocity dispersion of the system.
Observational data are currently being collected on merger rates in different
environments \citep{dokkum,patton}. In future work, we plan to carry out a 
more direct comparison with observational data of this kind, and so to test the
hierarchical paradigm for galaxy formation in new ways.

\section*{Acknowledgements}

The simulations presented in this paper were carried out on the T3E
supercomputer at the Computing Center of the Max-Planck-Society in
Garching, Germany and on the IBM SP2 at CINES in Montpellier, France.  \\
We thank Antonaldo Diaferio, Gao Liang and the referee, Fabio Governato, for
useful comments and discussions that significantly improved the 
presentation.\\ 
G.~D.~L. thanks the Alexander von Humboldt Foundation, the Federal
Ministry of Education and Research, and the Programme for Investment
in the Future (ZIP) of the German Government for financial support.

\bsp

\label{lastpage}

\bibliographystyle{mnras}
\bibliography{DeLucia_subs}

\end{document}